\documentclass[amsmath,amssymb,10pt,aps,notitlepage,prl,twocolumn,superscriptaddress,nofootinbib,longbibliography]{revtex4-1} 
 \usepackage[colorlinks=true]{hyperref}
 \usepackage{lipsum}
 \usepackage{graphicx}
\usepackage{latexsym}
\usepackage{amsmath}
\usepackage{amsthm}
\usepackage{amssymb}
\usepackage{epstopdf} 
\usepackage{enumitem}
\usepackage{setspace}
\usepackage{dcolumn}
\usepackage{bm}
\usepackage{setspace} 
\usepackage{slashed}
\usepackage{color}
\usepackage{youngtab}
\usepackage{tikz}
\usepackage{tikz-3dplot}
\usepackage{braket}
\usepackage{cancel}
\usepackage{subfigure}

\usetikzlibrary{shapes,snakes,arrows,chains,matrix,positioning,scopes,calc}
\usetikzlibrary{decorations.markings}
\usepackage{soul}

 \usepackage{moreverb}
\allowdisplaybreaks 
\begin{document}

\title{  Lattice vibration as a knob for novel quantum criticality  \\ : Emergence of supersymmetry from spin-lattice coupling  }
  
\author{SangEun Han}
\thanks{These authors contributed equally to this work. }
\affiliation{Department of Physics, Korea Advanced Institute of Science and Technology, Daejeon 34141, Korea}
\author{Junhyun Lee}
\thanks{These authors contributed equally to this work. }
\affiliation{Department of Physics, Condensed Matter Theory Center and the Joint Quantum Institute, University of Maryland, College Park, Maryland 20742, USA}
\author{Eun-Gook Moon}
\thanks{egmoon@kaist.ac.kr}
\affiliation{Department of Physics, Korea Advanced Institute of Science and Technology, Daejeon 34141, Korea}

\date{\today}
\begin{abstract}
Control of quantum coherence in many-body system is one of the key issues in modern condensed matter. Conventional wisdom is that lattice vibration is an innate source of decoherence, and amounts of research have been conducted to eliminate lattice effects.  
Challenging this wisdom, here we show that lattice vibration may not be a decoherence source but an impetus of a novel coherent quantum many-body state. 
We demonstrate the possibility by studying the transverse-field Ising model on a chain with renormalization group and density-matrix renormalization group method, and theoretically discover  a stable $\mathcal{N}=1$ supersymmetric quantum criticality with central charge $c=3/2$.
Thus, we propose an Ising spin chain with strong spin-lattice coupling as a candidate to observe supersymmetry. 
Generic precursor conditions of novel quantum criticality are obtained by generalizing the Larkin-Pikin criterion of thermal transitions. Our work provides a new perspective that lattice vibration may be a knob for exotic quantum many-body states.
 \end{abstract}

\maketitle

Quantum states on a lattice inevitably couple to lattice vibration, and the coupling is known to be one of the main sources of decoherence of quantum states. 
To be specific, let us consider a spin system on a lattice whose Hamiltonian may be schematically written by,
\begin{align*}
H = H_{\text{spin}}^{0} + H_{\text{phonon}}^{0} + g H_{s-l}, 
\end{align*}
where $H_{\text{spin}/\text{phonon}}^{0}$ is for a pure spin/phonon system, and $H_{s-l}$ is for the spin-lattice coupling. 
We introduce a dimensionless coupling constant ($g$) to characterize the strength of the coupling. 

Phonons usually play the role of an environment, and a spin quantum state becomes decoherent due to the spin-lattice coupling. 
In other words, a disentangled quantum state ($| \Psi_{\text{spin}} \rangle \otimes |\Psi_{\text{phonon}} \rangle$) is generically not an eigenstate of the total Hamiltonian. 
For example, magnon excitations may decay into acoustic phonons \cite{Kittel}, and spin qubits may develop spin-relaxation time \cite{Awschalom}. 
In terms of the spin-lattice model, the Hamiltonian without the spin-lattice coupling may be expressed by $
H_{\text{spin}}^{0} + H_{\text{lattice}}^{0} \simeq \sum_{\alpha} E_{\alpha} | \alpha \rangle \langle \alpha | + \sum_{q} \omega_q b_q^{\dagger} b_q$, 
where an eigenenergy ($E_{\alpha}$) with a quantum number ($\alpha$) of spins, acoustic phonon energy spectrum ($\omega_q$), and phonon creation/annihilation operators ($b_{q}^{\dagger}/ b_q$) with momentum $q$ are introduced.  
The ground state with characteristic length/time scales such as an excitation energy gap of quasi-particle excitations \cite{sachdev2011} loses coherence by coupling with acoustic phonons, as manifested in the perturbative calculation, 
\begin{equation}
\begin{aligned}
|G_{\text{spin}} \rangle \simeq&\; |G_{\text{spin}}^{0} \rangle \\
&\;+ g \sum_{\beta, \{b \}} | \beta ; \{b_q \}\rangle \frac{ \langle \beta ; \{b_q \}| H_{s-l} | \alpha; \{b_p \} \rangle}{E_G^{0} - E_{\beta; \{b_q \} }} +\cdots,
\end{aligned} \label{eq:perturb_state}
\end{equation}
where disentangled excited states $| \beta ; \{b_q \}\rangle$ are used. 
With the energy gap ($\Delta > 0$), the denominator of the second term may be safely approximated as $  E_{\beta; \{b_q \}}-E_G^{0} \gtrsim \Delta$, and it is apparent that the spin-lattice coupling acts as a decoherence source of a pure quantum state of spins.   
The relaxation and decay rates are estimated as $\tau^{-1} \propto g^2 $ for small $g$. 
It is widely believed that elimination of the lattice coupling is crucial to control coherence of quantum many-body states \cite{Xu}. 

\begin{figure}[tb]
\centering
\includegraphics[width=0.8\linewidth]{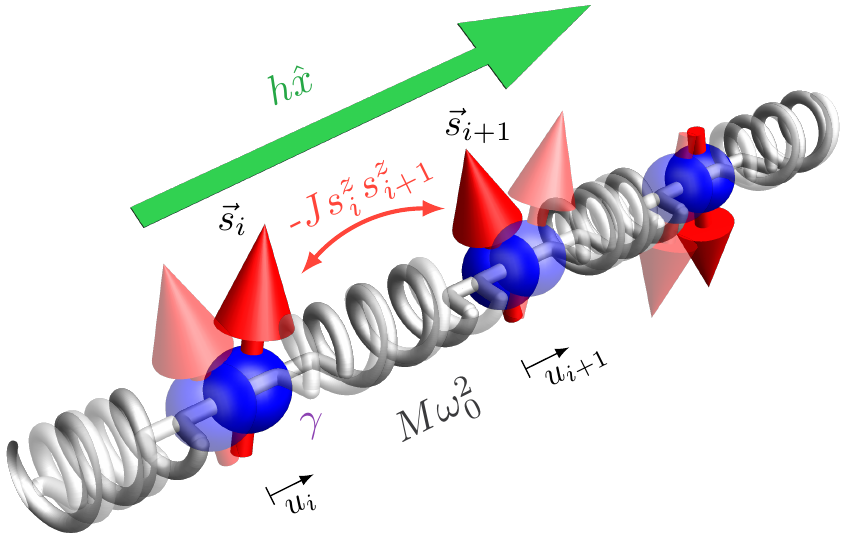}
\caption{\textbf{Illustration of the transverse Ising model under lattice vibrations}. The red arrow stands for the Ising spin and the springs represent the vibrating lattices. The green arrow at the top shows the transverse field and $\gamma$ is the coupling constant between the spin degree of freedom and lattice vibrations.\\
}\label{fig:scheme_isinglattice}
\end{figure}

Here, we challenge the common belief by demonstrating that lattice vibration may  be used to realize a novel coherent quantum many-body state. 
Especially, for a quantum critical state,  the above perturbative discussion is invalid due to gapless critical excitations, and instead, lattice vibration opens up the possibility of a novel quantum critical state by entangling lattice vibration and quantum critical modes as we show below. 
In other words, spin-lattice coupling may strongly drive a system to be in a pure state as a whole.

For a proof of principle, we consider the transverse-field Ising chain model with acoustic phonons. The Hamiltonian without spin-lattice coupling is
\begin{align}
H_{0}=&\sum_{i} \Big[ - J  s_{i}^{z} s_{i+1}^{z} -h s_{i}^{x} +\frac{P_{i}^{2}}{2M} + \frac{M \omega_{0}^{2}}{2} (u_{i+1}- u_{i} )^{2} \Big], \label{eq:H0} 
 \end{align}  
with a magnetic exchange interaction $J$, a transverse magnetic field $h$, Debye frequencey $\omega_0$, and ion mass $M$ (Fig.\ref{fig:scheme_isinglattice}). 
The deviation of spin positions are captured by $u_i$ and the quantum spins are represented by the Pauli matrices ($s_{j}^{x,y,z}$) at site $j$. 
The Hamiltonian is exactly solvable and becomes $
H_0= \sum_k \epsilon_k (f_k^{\dagger} f_k-\frac{1}{2} ) + \omega_k ( b_k^{\dagger} b_k+\frac{1}{2})$.
The bosonic operators ($b_k, b_k^{\dagger}$) describe acoustic phonons with the energy spectrum, $\omega_k = 2\omega_0 |\sin(\frac{k a}{2})|$, and the fermionic ones ($f_k, f_k^{\dagger}$) are from the Jordan-Wigner transformation of spins and have energy spectrum of $\epsilon_k = 2J \sqrt{ 1-2r \cos(k a) +r^2 }$. 
Lattice spacing $a$ and the ratio $r = h/J$ are introduced.  
Note that the pure spin term may also be represented by two Majorana fermions at each site ($\eta_j^{(1,2)}$). For example, the spin exchange term becomes $ s_j^z s_{j+1}^z =  -i \eta_{j}^{(2)}\eta_{j+1}^{(1)}$ in this representation. 
At $r=1$, gapless Majorana fermion excitation arises in the pure spin model, indicating the Ising universality class of central charge $c=1/2$ \cite{sachdev2011}.
On the other hand, the phonon spectrum is gapless because phonons are Goldstone bosons of translational symmetry. 

The spin-lattice coupling appears with spatial modulation of the magnetic exchange interaction, 
$J \rightarrow J_{i,i+1}= J + \gamma (u_{i+1} - u_i) + O((u_{i+1} - u_i)^2)$ \cite{doi:10.1143/JPSJ.53.1472}, and the leading interaction term is
\begin{align}
H_1 = \gamma \sum_i (u_{i+1}-u_i)s_i^z s_{i+1}^z. \label{eq:H1}
\end{align}
Away from the critical point ($r \neq 1$), perturbative calculation indicates that decay rate of a quantum state is indeed proportional to $\tau^{-1} \propto \gamma^2$, and the spin-lattice coupling becomes a source of decoherence.

\begin{figure}[tb]
\centering
\includegraphics[width=0.8\linewidth]{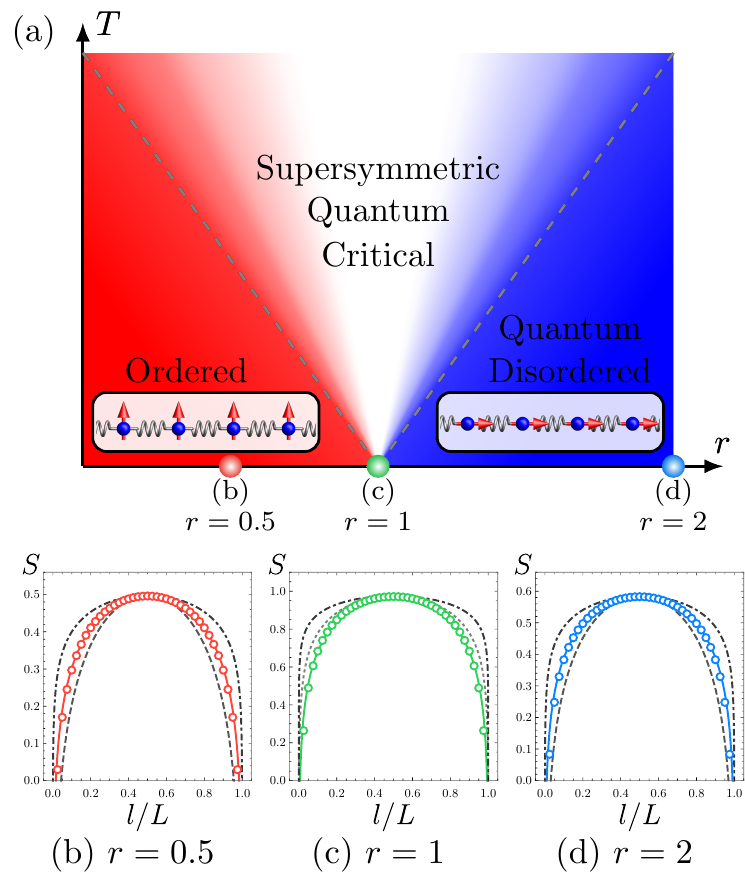}
\caption{\textbf{Phase diagram and entanglement entropy plot for the transverse Ising model under lattice vibrations}. (a) Phase diagram of the transverse Ising model under lattice vibrations. $r<1$ corresponds to the ordered phase and the spin degrees of freedom are aligned along the $z$ axis, while $r>1$ indicates the system is in the quantum disordered phase and the spin degrees of freedom are along to the $x$ axis. $r=1$ and $T=0$ is the quantum critical point which is described by the $\mathcal{N}=1$ supersymmetric conformal field theory (CFT) with central charge $c=3/2$. (b, c, d) DMRG calculation of the entanglement entropy of the system for three different values of $r$ as indicated in the phase diagram (a). The $x$ axis is the location of the boundary of the two subsystems ($L$ is the length of the system while $l$ is that of the subsystem). CFT predicts the scaling of entanglement entropy and the result for central charge $c=1/2$ (dash-dotted line), $c=1$ (dotted line), and $c=3/2$ (dashed line) are plotted as a comparison. At the critical point [(c)], the scaling suggests central charge of $3/2$ while away from the critical point [(b,d)], central charge is $1$. The fitted CFT scaling is shown as a solid line in all three figures. }\label{fig:phase_diagram}
\end{figure}

\begin{figure}[tb]
\centering
\subfigure[]{
\includegraphics[height=0.55\linewidth]{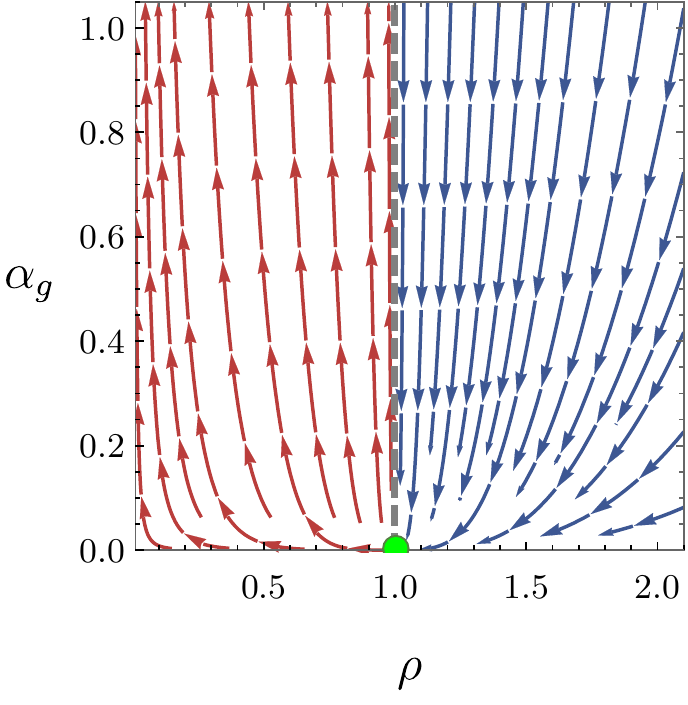}\label{fig:rg_flow}
}\\
\subfigure[]{
\includegraphics[height=0.315\linewidth]{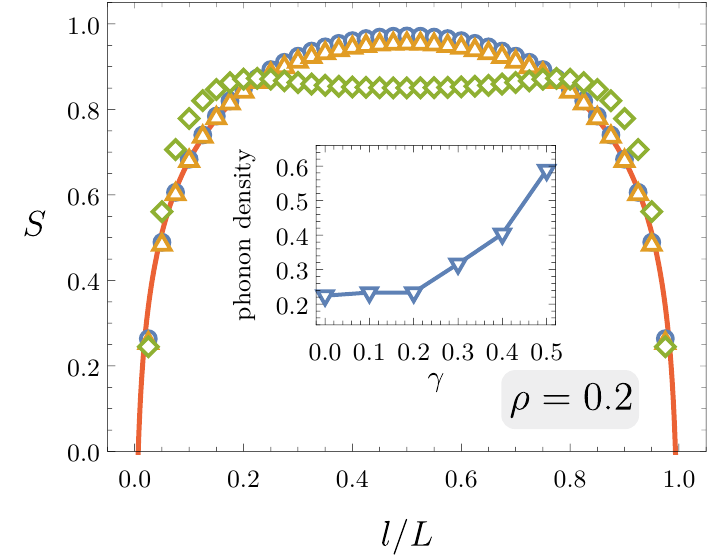}\label{fig:rho02}
}
\subfigure[]{
\includegraphics[height=0.315\linewidth]{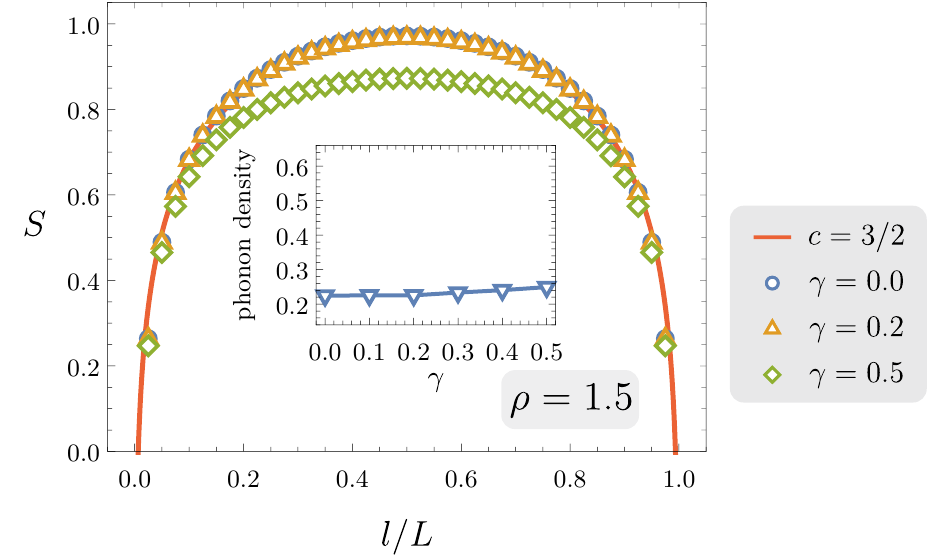}\label{fig:rho15}
}
\caption{{\bf RG flow diagram with the two dimensionless parameters, the velocity ratio of phonons and spinons $\rho$ ($\equiv v_{s}/v_{M}$) and the spin-lattice coupling constant $\alpha_g$ ($\equiv g^2 / (2\pi v_{s}^2 v_{M})$), and bipartite entanglement entropy plots in terms of $\rho$ and $\gamma$.} (a) While the RG flow (red) is directed to $(\rho,\alpha_{g})=(0,\infty)$  for $\rho<1$, the flow (blue) is directed to\ $(\rho,\alpha_{g})=(1,0)$,  for $\rho>1$. The fixed point is described by $\mathcal{N}=1$ superconformal field theory with central charge $c=3/2$. (b, c) DMRG calculations of the entanglement entropy with spin-lattice coupling $\gamma$. The two plots represents each side ($\rho<1$ and $\rho>1$) of the flow diagram as $\rho = 0.2$ for (b) and $\rho = 1.5$ for (c). One can observe the strong deviation from the original CFT for $\rho<1$. The inset shows how the average phonon occupancy changes with $\gamma$. The significant increase of phonon density in $\rho<1$ indicates that the phonons are responsible for the state flowing away from the decoupled theory. }\label{fig:flow_diagram}
\end{figure}

Now, let us consider a quantum critical state.  The scale invariance allows us to use the critical theory of spin and lattice degrees of freedom, whose form  is,
\begin{align*}
\mathcal{S}_{0} =& \int_{\tau, x}\;\frac{1}{2}\eta^{\intercal}\left( \partial_{\tau}+iv_M \sigma_{x}\partial_{x}\right)\eta + \frac{1}{2} (\partial_{\tau} u)^{2}+\frac{v_s^2}{2} (\partial_x u)^{2},
 \end{align*}
where the Pauli matrices ($\sigma_{x,y,z}$) are defined in the two component Majorana spinor $\eta^{\intercal} = (\eta^{(1)} \,\,\, \eta^{(2)})$ space. 
The Majorana fields are rescaled to have the factor $1/2$, and the short-handed notation $\int_{\tau,x} \equiv \int d \tau d x$ is used hereafter. 
 The two velocities ($v_M= 2J a$, $v_s = \omega_0 a $) are associated with magnetic exchange and Debye energy scales, respectively.
 Then, the spin-lattice coupling may be identified as  
\begin{align*}
\mathcal{S}_1 =  \frac{g}{2} \int_{\tau,x} ( \partial_{x}u)  \eta^{T}\sigma_{y}\eta. \nonumber
\end{align*}
The total critical theory, $\mathcal{S}_0 +  \mathcal{S}_1$, are analyzed by introducing the two dimensionless coupling constants, 
$\rho \equiv v_s/v_M$ and $\alpha_{g}\equiv g^{2}/ (2\pi v_s^{2}v_M) $.
We perform the one loop renormalization group (RG) analysis and obtain the flow equations,
\begin{align}
 \frac{d\rho}{d\ell}=-\frac{\alpha_{g}}{2}  \left(\frac{1-\rho}{1+\rho}\right)^{2} \rho,\quad \frac{d\alpha_{g}}{d\ell}={\alpha_{g}^2}\left(\frac{1-\rho}{1+\rho}\right).\label{eq:rgflow}
\end{align}
In Fig.~\ref{fig:rg_flow}, the flow diagram is illustrated, and the fixed point is at $(\rho,\alpha_{g})=(1,0)$. 
The RG flow around the fixed point is intriguing.  
 If the phonon is slower than the Majorana fermion ($\rho <1$), the flow goes to $(\rho,\alpha_{g})=(0,\infty)$ signaling a first order phase transition.  
In the opposite case where  the phonon is faster than the Majorana fermion ($\rho >1$), the RG flow is directed to the stable fixed point, $(\rho,\alpha_{g})=(1,0)$. 

The Hamiltonian at the stable fixed point may be written as 
\begin{align}
H_{\text{sc}} = J \sum_i \Big[ -  s_i^z s_{i+1}^z - s_i^x +p_i^2 + (x_{i+1}- x_i )^2 \Big],
\end{align}
with rescaled momentum and position operators, $p_j \equiv P_j/\sqrt{M \omega_0}$ and $x_j \equiv \sqrt{M \omega_0} u_j $.
We introduce an operator, 
\begin{align}
\hat{Q} = -\sum_j \big( \prod_{l<j} s_l^x \big) \big[ (x_{j-1}-x_j) s^z_j +  p_j s_j^y \big],
\end{align}
 which is fermionic, proven by the Jordan-Wigner transformation, and satisfies $ H_{\text{sc}} = J \hat{Q}^2 $. Therefore, this fermionic operator becomes a supercharge, $[\hat{Q}, H_{sc}]=0$, and the $\mathcal{N}=1$ supersymmetry with central charge $c=3/2$ is obtained.

Our analysis is also checked by the density-matrix renormalization group (DMRG) method. 
For a chain of length $L=40$, we calculate the ground state of the system and its bipartite entanglement entropy for different ratios of energy scales ($J, h, \gamma$). 
The central charge can be obtained from the entanglement entropy scaling with the subsystem size. 
Without spin-lattice coupling, we find $c=3/2$ at the critical point ($r=1$) while $c=1$ otherwise as the spin sector becomes gapped and only the acoustic phonon contributes (Fig.~\ref{fig:phase_diagram}(b)(c)(d)). 
When the interaction is turned on and the system is at the critical point, two distinct behaviors arise for regions $\omega_0 > 2 J$ and $\omega_0 < 2 J$. 
The system is stable for $\omega_0 > 2 J$, manifested by the central charge unchanged at $c=3/2$, while big deviations from $c=3/2$ occurs for $\omega_0 < 2 J$, which is consistent with the results of the RG flow (Fig.~\ref{fig:rho02} and \ref{fig:rho15}).
We have also measured the phonon density in each calculation and observed that the phonon occupancy significantly increases for nonzero coupling and $\omega_0 < 2 J$, supporting the conventional decoherence from phonons in that parameter regime.

Few comments are as follows. 
First, the supersymmetric quantum state emerges from the spin-lattice coupling. 
Without the coupling, the state loses supersymmetry unless interactions are fine-tuned. 
Note that the interactions from spin-lattice are unique in a sense that bosons have a shift symmetry, $u_i \rightarrow u_i +a$, in contrast to ladder systems \cite{PhysRevLett.114.090404,Garst, PhysRevB.87.165145}.
Second, the supersymmetric quantum criticality cannot be obtained by the standard quantum-classical mapping. 
It is because lattice vibrations are intrinsically tied to spatial dimensions. 
 To illustrate this, we consider the classical Ising model in 2d, $H_{\text{Ising}} = - J \sum_{\langle i,j \rangle} \sigma_i^z \sigma_j^z$, which may be mapped to the transverse-field Ising chain model \cite{sachdev2011}. The 2d static phonon Hamiltonian in cubic solids may be described by $H_{\text{ph}}= \int d^{2}x\;\frac{1}{2}\left[C_{11}(e_{11}^{2}+e_{22}^{2})+2C_{12}e_{11}e_{22}+C_{44}e_{12}^{2}\right]$,
where $e_{ab}\equiv\frac{1}{2}(\partial_{a}u_{b}+\partial_{b}u_{a})$ is the strain field, $u_{a}$ is the acoustic phonon field with $a=1,2$, and $C_{ij}$'s are the elastic stiffness constants \cite{kittel1996introduction}.
 There are two phonon modes along the two spatial directions in sharp contrast to the one mode in the quantum model, and moreover, the bulk modulus, $K = (C_{11}+C_{12})/2$, keeps decreasing under the scale transformation, which indicates the instability of the thermal Ising transition. 
Thus, the $\mathcal{N}=1$ supersymmetry at the critical point cannot appear in the thermal transition.  
Third, the origin of the supersymmetry in our work is different from the previously suggested ones in the literature \cite{PhysRevB.87.165145,PhysRevLett.114.090404, SSLee, Grover, Ponte, Yao, PhysRevLett.116.100402,PhysRevLett.118.166802} where bosons are made of fermions, and special types of interactions or surface degrees of freedom are necessary. 
In contrast, bosons are from the lattice and fermions are from the spins in this work. 
One crucial point is that bosons are in a critical phase in a sense that its spectrum is always gapless, and spin degrees of freedom realize gapless fermionic excitations at the critical point.

Going beyond the spin chain system, let us consider a generic Landau-Ginzburg Hamiltonian with a local order parameter $\phi$, 
\begin{align}
H= - \sum_{\langle ij \rangle} t \, \phi_i \phi_j + \sum_{ijkl} \lambda \, \phi_i \phi_j \phi_k \phi_l+ H_{\text{ph}}.  
 \end{align}
Here the indices are for the positions of the order parameters. 
For simplicity, we consider the case where the symmetry group of the order parameter is decoupled from that of the lattice in this work, leaving other cases for future works. 
The lattice Hamiltonian $H_{\text{ph}}$ generally consists of harmonic and anharmonic terms with an additional polarization index.  
As in the Ising model, we promote the coupling at the lowest order of $\phi$ to have a spatial modulation to introduce minimal interaction between the order parameter and the lattice vibrations: $t \rightarrow t_{ij}= t + \gamma (u_j - u_i) + O((u_j - u_i)^2)$.

A quantum phase transition may be described by tuning the parameter $\lambda/t$.
To study the behavior near the phase transition, we again consider the critical field theory. 
The total action $\mathcal{S} = \mathcal{S}_{c}+ \mathcal{S}_{\text{ph}}+\mathcal{S}_{\text{ph-}c}$ describes the interplay physics between quantum criticality and acoustic phonons, where $\mathcal{S}_{c}$ is the critical action for the original theory for the order parameter, $\mathcal{S}_{\text{ph}}$ is the action for the acoustic phonons, and $\mathcal{S}_{\text{ph-}c}$ represents the interaction of the two. 
The coupling term in the action, $\mathcal{S}_{\text{ph-}c}$, is solely determined by the symmetry of the theory. 
Since we are considering the case where the order parameter and the lattice represents different symmetries, the most relevant interaction term is
$\mathcal{S}_{\text{ph-}c}= g \int_{\tau,x} \mathcal{O}_{E} \sum_{i=1}^{d} \partial_i u_i $.
$u_i$ is the phonon field and $d$ is the number of spatial dimensions. 
The form of the energy operator $\mathcal{O}_E$ depends on the system, for example, $\mathcal{O}_{E}=\phi^2/2$ in the conventional $\phi^4$ theory,
$\mathcal{S}_{c}=\int_{\tau, x} \frac{1}{2}\left((\partial_{\tau} {\phi})^{2}+v^{2}(\nabla {\phi})^{2}\right)+\frac{r}{2} {\phi}^{2}
+\frac{\lambda}{4!}{\phi}^{4}$.

The standard scaling analysis may be performed at the fixed point without the lattice-order parameter coupling.
The strain tensor and the energy operator have $[e_{ij}]=\frac{d+z}{2}$ and $ [\mathcal{O}_E]=z+d -\frac{1}{\nu}$. The scaling dimension of $g$ is 
\begin{align} 
[g]=\frac{1}{\nu}-\frac{d+z}{2}=\frac{2-(d+z)\nu}{2\nu},  
\end{align}
whose sign becomes the main criterion for the stability. 
For $[g] <0$, the quantum criticality of $\mathcal{S}_c$ is stable, so the ground state may be described by a disentangled state of order parameters and phonons. 
Perturbative calculations give rise to decoherence of quantum states of order parameters. 
But, for $[g]\geq0$, the disentangled state becomes unstable indicating two possibilities. First, the second-order phase transition may become a first-order transition as in most thermal phase transitions under lattice vibration. Second, as in the above spin-chain model, a novel quantum criticality may appear.  
As an example, the scaling dimensions of the lattice-order parameter in several models which have $z=1$ are presented in Table.~\ref{tab:stab_crit}.
Note that our condition becomes the Larkin-Pikin criterion \cite{larkin1969phase} in the limit of classical phase transitions. Namely, setting $z=0$, $[g]<0$ becomes the negative heat capacity critical exponent $\alpha=2 -d \nu <0$, and the corresponding classical criticality is stable.
 
 \begin{table}[t]
\begin{tabular}{|c|c|>{$}c<{$}|>{$}c<{$}|}
\hline
Model&  \quad$d_{\text{eff}}$ \quad &\nu&\quad [g] \quad \quad \\
\hline\hline
Ising \cite{DiFrancesco1997-7}&2&1&0 \\
Tricritical Ising  \cite{DiFrancesco1997-7}&2& 5/9& >0 	\\
3 (4)-state Potts \cite{DiFrancesco1997-7}&2&5/6 \, (2/3)&>0	\\
$q$-state clock ($q>4$) \cite{Tobochinik}&2&\infty& <0 	\\
\hline \hline
Ising \cite{Vicari_RG}&3&0.63&>0	\\
$q$-state clock ($q\geq4$)\cite{Vicari}&3&0.67&<0\\
O($N$)  ($N\geq2$) \cite{Vicari,Vicari_RG}&3&\geq0.67&<0\\
$\mathcal{N}=2$ WZ SUSY \cite{PhysRevLett.116.100402}&3& 0.917 & <0 \\
$\mathcal{N}=2$ XYZ SUSY  \cite{PhysRevLett.118.166802}&3&1/2+ \epsilon/4& \,\,<0  \\
\hline \hline
O($N$) \cite{sachdev2011}&4&1/2&0 \\
\hline
\end{tabular}
\caption{Scaling dimensions of the coupling ($[g]= 1/\nu- d_{\text{eff}}/2$) in models with continuous phase transitions. The effective dimension $d_{\text{eff}}\equiv(d+1)$ is introduced for spatial dimension $d$.  For $[g]>0$, the original criticalities become unstable signaling the first-order phase transition under lattice vibration, and for $[g]<0$, the original criticalties are stable  under lattice vibrations. For $[g]=0$, a novel quantum criticality may appear as in the spin-chain model in the main text.}\label{tab:stab_crit}
\end{table}

The generalized Larkin-Pikin criterion may be also applied to unconventional quantum criticalities. 
First, topological phase transitions in weakly correlated systems are generically described by the Dirac/Weyl fermions, whose Hamiltonian is written as $H_{D/W} = \int d^d x \psi^{\dagger} ( -i \partial_{a} \Gamma^a)\psi$ with $a=1,\cdots, d$, the Clifford algebra matrices $\Gamma_a$, and the spinor $\psi$ \cite{RevModPhys.90.015001}. The sign of the mass determines whether the system is in the topological phase, and the correlation length critical exponent is $\nu=1$. Setting $z=1$, the coupling constant is marginal in $d=1$ and irrelevant for $d>1$. 
For criticalities with $z>1$, the coupling becomes less irrelevant, but is still irrelevant at higher dimensions such as $d=3$.
Thus, new universality class or instability may appear at $d=1$ while topological phase transitions in higher dimensions may be decoupled from the lattice vibration. 
Second, quantum criticalities with an enlarged symmetry, such as criticalities in a deconfined phase, may have different universality class from that of the Landau-Ginzburg-Wilson paradigm \cite{Senthil1490,PhysRevX.7.031051}. For example, a $Z_2$ symmetry breaking transition with $Z_2$ local gauge in $d=2$ has the same universality class as the one of $U(1)$ symmetry transition \cite{Sachdev_Topo, 2018arXiv181205621M}, so the lattice vibration becomes decoupled.  
Third, the criterion may be applied to the recently proposed quantum annealed criticality \cite{2018arXiv180511771C}, which connects a quantum critical point with a line of first-order thermal phase transitions. 
One good candidate is the $Z_4$ clock model in $d=2$. At zero temperature the model shows the $U(1)$ universality class because a four-fold anisotropy is reported to be irrelevant \cite{Vicari_RG, Vicari}, but at non-zero temperatures, the model shows non-universal behaviors \cite{Taroni_2008}. Namely, its universality class may be the same as one of the Ashkin-Teller (4-state Potts) model with $\nu_{AT} =2/3$ depending on systems' parameters \cite{Kadanoff}, indicating a first-order transition.  
Lastly, the criterion indicates that the interplay between lattice vibration and quantum criticality may be accessed perturbatively in recently reported ferroelectric quantum criticalities in SrTiO$_3$ and KTaO$_3$ \cite{Rowley2014, STO_SC,THz1,THz2}.

Our results provide non-trivial predictions in experiments of emergent phenomena in quantum material. The $\mathcal{N}=1$ supersymmetry in the spin chain model indicates that the velocity of acoustic phonons becomes equalized to the spinon velocity, which may be tested by sound attenuation experiments, for example in CoNb$_2$O$_6$ \cite{Coldea177}. 
The phonon velocity is generically faster than the spinon velocity, so we predict significant decreases of phonon velocity around the quantum critical point and the two eventually become equal in the ideal case. 
Furthermore, the coupling constant of the interplay physics in three spatial dimensions is marginal at the tree level, and thus logarithmical corrections are expected in physical quantities which will be discussed in future works.

In conclusion, we demonstrate that lattice vibration may be an impetus of a novel quantum many-body state, not an intrinsic source of decoherence. 
A whole system with spin and lattice degrees of freedom may form a macroscopic quantum many-body state by entangling quantum critical modes and acoustic phonons. One example we discover in this work is a supersymmetric quantum criticality of an Ising spin-chain. Its striking characteristics of the entanglement may be observed in experiments, for example, equal phonon and spinon velocities in the Ising chain. 
Our results indicate that interplay between quantum criticality and lattice vibration may open a new regime of quantum many-body physics.

\textit{Acknowledgement} :
 We thank P. Coleman, H. Katsura, and S. S. Lee for invaluable discussions and comments.    
This work was supported by NRF of Korea under Grant No. 2017R1C1B2009176 (SH, EGM), and NSF-PFC at the JQI (JL).

%
%
%

\end{document}